\documentclass[aps,amsmath,amssymb,twocolumn,superscriptaddress,nofootinbib]{revtex4-1}

\usepackage{graphicx}
\usepackage{dcolumn}
\usepackage{bm}
\usepackage{color}
\usepackage[colorlinks=true,pdfstartview=FitV,linkcolor=blue,citecolor=blue,urlcolor=blue]{hyperref}

\everymath{\displaystyle}
\begin{document}

\newcommand{\up}[1]{$^{#1}$}
\newcommand{\down}[1]{$_{#1}$}
\newcommand{\powero}[1]{\mbox{10$^{#1}$}}
\newcommand{\powert}[2]{\mbox{#2$\times$10$^{#1}$}}

\newcommand{\evm}{\mbox{eV\,$c^{-2}$}}
\newcommand{\pgd}{\mbox{g$^{-1}$\,d$^{-1}$}}
\newcommand{\mv}{\mbox{$m_V$}}
\newcommand{\um}{\mbox{$\mu$m}}
\newcommand{\spix}{\mbox{$\sigma_{\rm pix}$}}
\newcommand{\pav}{\mbox{$\langle p \rangle$}}

\title{First Direct-Detection Constraints on eV-Scale Hidden-Photon Dark Matter\\ with DAMIC at SNOLAB}

\author{A.~Aguilar-Arevalo} 
\affiliation{Universidad Nacional Aut{\'o}noma de M{\'e}xico, Mexico City, Mexico} 

\author{D.~Amidei}
\affiliation{Department of Physics, University of Michigan, Ann Arbor, MI, United States}  

\author{X.~Bertou}
\affiliation{Centro At\'omico Bariloche, CNEA/CONICET/IB, Bariloche, Argentina}

\author{M.~Butner}
\affiliation{Fermi National Accelerator Laboratory, Batavia, IL, United States}
\affiliation{Northern Illinois University, DeKalb, IL, United States }

\author{G.~Cancelo}
\affiliation{Fermi National Accelerator Laboratory, Batavia, IL, United States}

\author{A.~Casta\~{n}eda V\'{a}zquez}
\affiliation{Universidad Nacional Aut{\'o}noma de M{\'e}xico,  Mexico City, Mexico} 

\author{B.A.~Cervantes Vergara}
\affiliation{Universidad Nacional Aut{\'o}noma de M{\'e}xico, Mexico City, Mexico} 

\author{A.E.~Chavarria}
\affiliation{Kavli Institute for Cosmological Physics and The Enrico Fermi Institute, The University of Chicago, Chicago, IL, United States}

\author{C.R.~Chavez}
\affiliation{Facultad de Ingenier\'{\i}a, Universidad Nacional de Asunci\'on, Asuncion, Paraguay}

\author{J.R.T.~de~Mello~Neto}
\affiliation{Universidade Federal do Rio de Janeiro, Instituto de  F\'{\i}sica, Rio de Janeiro, Brazil}

\author{ J.C.~D'Olivo}
\affiliation{Universidad Nacional Aut{\'o}noma de M{\'e}xico, Mexico City, Mexico} 

\author{J.~Estrada}
\affiliation{Fermi National Accelerator Laboratory, Batavia, IL, United States}

\author{G.~Fernandez~Moroni}
\affiliation{Fermi National Accelerator Laboratory, Batavia, IL, United States}
\affiliation{Universidad Nacional del Sur, Bahia Blanca, Argentina}

\author{R.~Ga\"ior}
\affiliation{Laboratoire de Physique Nucl\'eaire et de Hautes Energies (LPNHE), Universit\'es Paris 6 et Paris 7, CNRS-IN2P3, Paris, France}

\author{Y.~Guardincerri}
\affiliation{Fermi National Accelerator Laboratory, Batavia, IL, United States}

\author{ K.P.~Hern\'{a}ndez~Torres}
\affiliation{Universidad Nacional Aut{\'o}noma de M{\'e}xico, Mexico City, Mexico} 

\author{F.~Izraelevitch}
\affiliation{Fermi National Accelerator Laboratory, Batavia, IL, United States}

\author{A.~Kavner}
\affiliation{Department of Physics, University of Michigan, Ann Arbor, MI, United States}  

\author{B.~Kilminster}
\affiliation{Universit{\"a}t Z{\"u}rich Physik Institut, Zurich, Switzerland }

\author{I.~Lawson}
\affiliation{SNOLAB, Lively, ON, Canada }

\author{A.~Letessier-Selvon}
\affiliation{Laboratoire de Physique Nucl\'eaire et de Hautes Energies (LPNHE), Universit\'es Paris 6 et Paris 7, CNRS-IN2P3, Paris, France}

\author{J.~Liao}
\affiliation{Universit{\"a}t Z{\"u}rich Physik Institut, Zurich, Switzerland }

\author{A.~Matalon}
\affiliation{Kavli Institute for Cosmological Physics and The Enrico Fermi Institute, The University of Chicago, Chicago, IL, United States}

\author{V.B.B.~Mello}
\affiliation{Universidade Federal do Rio de Janeiro, Instituto de  F\'{\i}sica, Rio de Janeiro, Brazil}

\author{J.~Molina}
\affiliation{Facultad de Ingenier\'{\i}a, Universidad Nacional de Asunci\'on, Asuncion, Paraguay}

\author{P.~Privitera}
\affiliation{Kavli Institute for Cosmological Physics and The Enrico Fermi Institute, The University of Chicago, Chicago, IL, United States}

\author{K.~Ramanathan}
\affiliation{Kavli Institute for Cosmological Physics and The Enrico Fermi Institute, The University of Chicago, Chicago, IL, United States}

\author{Y.~Sarkis}
\affiliation{Universidad Nacional Aut{\'o}noma de M{\'e}xico, Mexico City, Mexico} 

\author{T.~Schwarz}
\affiliation{Department of Physics, University of Michigan, Ann Arbor, MI, United States}  

\author{M.~Settimo}
\affiliation{Laboratoire de Physique Nucl\'eaire et de Hautes Energies (LPNHE), Universit\'es Paris 6 et Paris 7, CNRS-IN2P3, Paris, France}

\author{M.~Sofo~Haro}
\affiliation{Centro At\'omico Bariloche, CNEA/CONICET/IB, Bariloche, Argentina}

\author{R.~Thomas}
\affiliation{Kavli Institute for Cosmological Physics and The Enrico Fermi Institute, The University of Chicago, Chicago, IL, United States}

\author{J.~Tiffenberg}
\affiliation{Fermi National Accelerator Laboratory, Batavia, IL, United States}

\author{E.~Tiouchichine}
\affiliation{Centro At\'omico Bariloche, CNEA/CONICET/IB, Bariloche, Argentina}

\author{D.~Torres Machado}
\affiliation{Universidade Federal do Rio de Janeiro, Instituto de  F\'{\i}sica, Rio de Janeiro, Brazil}

\author{F.~Trillaud}
\affiliation{Universidad Nacional Aut{\'o}noma de M{\'e}xico, Mexico City, Mexico} 

\author{X.~You}
\affiliation{Universidade Federal do Rio de Janeiro, Instituto de  F\'{\i}sica, Rio de Janeiro, Brazil}

\author{J.~Zhou}
\affiliation{Kavli Institute for Cosmological Physics and The Enrico Fermi Institute, The University of Chicago, Chicago, IL, United States}

\collaboration{DAMIC Collaboration}
\noaffiliation

\date{\today}

\begin{abstract}
We present direct detection constraints on the absorption of hidden-photon dark matter with particle masses in the range 1.2--30\,\evm\  with the DAMIC experiment at SNOLAB. Under the assumption that the local dark matter is entirely constituted of hidden photons, the sensitivity to the kinetic mixing parameter $\kappa$ is competitive with constraints from solar emission, reaching a minimum value of \powert{-14}{2.2} at 17\,\evm. These results are the most stringent direct detection constraints on hidden-photon dark matter in the galactic halo with masses 3--12\,\evm\ and the first demonstration of direct experimental sensitivity to ionization signals $<$12\,eV from dark matter interactions.
\end{abstract}


\maketitle

The DAMIC (Dark Matter in CCDs) experiment at SNOLAB~\cite{Aguilar-Arevalo:2016ndq} employs the bulk silicon of scientific-grade CCDs as a target for ionization signals produced by interactions of particle dark matter from the galactic halo. In this Letter, we report on a search for hidden photons, massive vector bosons that have been proposed as candidates to explain the origin of the dark matter in the Universe~\cite{Pospelov:2008jk, *Redondo:2008ec, *Nelson:2011sf, *Arias:2012az}.

Similarly to ordinary photons, hidden photons can be absorbed by electrons in the bulk of a silicon device and lead to an ionization signal. For the case of nonrelativistic hidden photons in the galactic halo, which release their rest energy in the target, free charge carriers may be produced for rest energies above the band gap of silicon (1.2\,eV), with a larger number of charge carriers produced with increasing hidden-photon mass ($m_V$). In this analysis, we consider ionization signals in the range 1--11\,$e^-$, probing the absorption of hidden photons with masses 1.2--30\,\evm .

The absorption cross section for hidden photons~\cite{An:2014twa, Hochberg:2016sqx, Bloch:2016sjj} is determined by the kinetic mixing $\kappa$ between the field strength tensors of electromagnetism and  its ``hidden" counterpart~\cite{Okun:1982xi, *Holdom:1985ag}. When in-medium dispersion effects are considered, an effective mixing parameter $\kappa_{\rm eff}$ can be defined such that the absorption cross section for a non-relativistic hidden photon 
in the medium, $\sigma_V(m_V)$, is related to the photoelectric cross section for a photon with energy $m_Vc^2$, $\sigma_{\gamma}(m_Vc^2)$, by
\begin{equation*} 
\sigma_V(m_V) v = \kappa^{2}_{\rm eff} \sigma_{\gamma}( \mbox{$m_Vc^2$} )c,
\end{equation*}
where $v$ is the speed of the hidden photon in the laboratory frame and $c$ is the speed of light.
The effective kinetic mixing can be expressed as
\begin{equation*} 
\kappa^{2}_{\rm eff} = \frac{\kappa^{2}m_{V}^{4}}{(m_{V}^{2}-{\rm Re}[\Pi(m_Vc^2)])^{2} + ({\rm Im}[\Pi(m_Vc^2)])^{2}},
\end{equation*}
where $\Pi(m_Vc^2)$ is the polarization tensor of the medium evaluated at a photon energy equal to the hidden photon rest energy.

Hence, for a detector target located within the dark matter halo the absorption rate of hidden photons would be
\begin{equation} 
\Gamma =  \frac{\rho_{\rm DM}}{m_{V}}\kappa_{\rm eff}^2\sigma_{\gamma}(m_Vc^2)c,
\label{eq:rate}
\end{equation}
where $\rho_{\rm DM}$$=$0.3\,GeV\,$c^{-2}$\,cm$^{-3}$ is the local density of dark matter. Of particular relevance is the lack of dependence of $\Gamma$ on $v$, the result being insensitive to the details of the hidden-photon velocity distribution in the dark matter halo. Thus, unlike searches for weakly-interacting massive particles by elastic scattering off nuclei, no annual modulation is expected in the potential signal.

 The search for hidden photons was performed with 6.25\,d of data acquired in January 2016 with a 4$k$$\times$4$k$-pixel CCD (5.8\,g in mass) deployed as part of the R\&D program of the DAMIC experiment.
 This device exhibited the lowest leakage current of four CCDs installed in the DAMIC copper box when cooled to 105$\pm$5\,K inside a copper vacuum vessel ($\sim$\powero{-6}\,mbar).
 The setup was shielded on all sides by at least 18\,cm of lead and 42\,cm of polyethylene to stop background radiation from environmental $\gamma$ rays and neutrons, respectively.
 Details of the DAMIC infrastructure at SNOLAB can be found in Refs.~\cite{Aguilar-Arevalo:2015lvd, Aguilar-Arevalo:2016ndq}.

\begin{figure}[b!]
\centering
\includegraphics[width=0.333\textwidth]{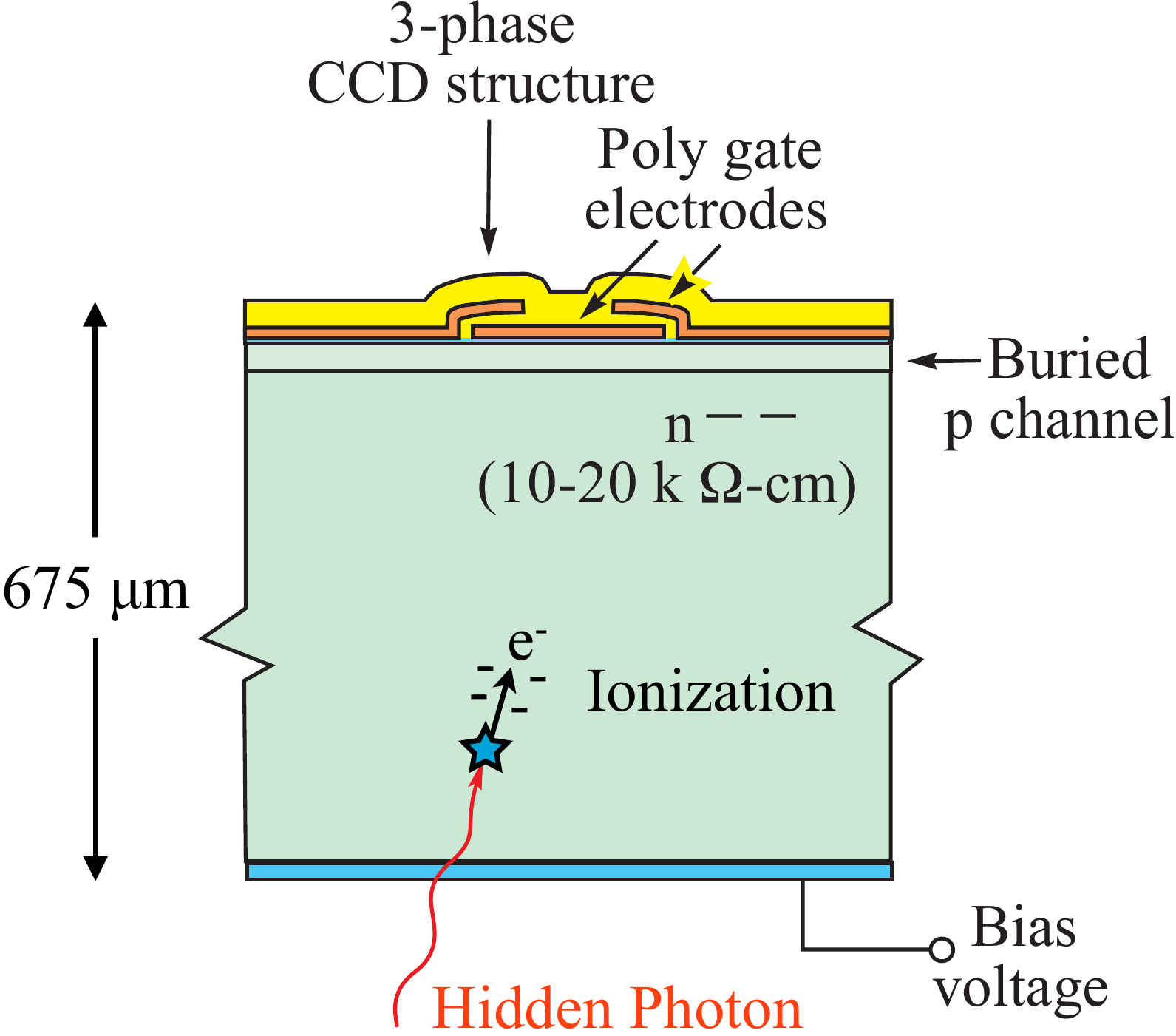}
\caption{Cross-sectional diagram of a pixel of a DAMIC CCD. A hidden photon is absorbed by a valence electron in the bulk silicon. The rest energy of the hidden photon ($m_Vc^2$) is released into an energetic photoelectron. The photoelectron loses its kinetic energy by ionization, generating secondary charge carriers in the silicon. The charge carriers are then drifted across the substrate by the applied electric field and held below the gates until the device is read out. Adapted from Ref.~\cite{1185186}.} \label{fig:pix}
\end{figure}

A CCD pixel consists of a three-phase polysilicon gate structure with a buried $p$ channel (Fig.~\ref{fig:pix}).
The pixel size is 15\,$\times$\,15\,\um$^{2}$ and the bulk of the device is high-resistivity (10--20 k$\Omega$\,cm) $n$-type silicon with a substrate thickness of 675\,\um.
The high resistivity of the silicon allows for a low donor density in the substrate ($\sim$\powero{11}\,cm\up{-3}), which leads to fully depleted operation at a substrate bias of 40\,V.
Charge produced in the bulk by ionization (e.g., from the absorption of a hidden photon, as in Fig.~\ref{fig:pix}) is drifted along the direction of the electric field across the substrate.
Because of thermal motion, the charge carriers diffuse transversely with respect to the electric field direction as they are drifted, with a lateral variance that is proportional to the carrier transit time.
The charge is collected and held near the $p$-$n$ junction, less than 1\,\um\ below the gates, until the device is read out.
During readout, the charge is transferred in the $y$ direction from pixel to pixel along each column by appropriate clocking of the three-phase gates (``parallel clocks''), while higher frequency clocks (``serial clocks'') move the charge of the last row (the ``serial register") in the $x$ direction to the primary CCD charge-to-voltage amplifier, the output node. A precise measurement of the charge is then performed by a correlated double-sampling circuit~\cite{janesick2001scientific}.
DAMIC CCDs have a second output node on the other end of the serial register in which charge is not deposited, offering a measurement of zero charge, i.e., of noise.
The inefficiency of charge transfer from pixel to pixel is as low as \powero{-6}~\cite{1185186} and the readout noise in the charge measurement is $\sim$2\,$e^-$~\cite{Aguilar-Arevalo:2016ndq}.
The image is reconstructed from the order in which the pixels are read out, and contains a two-dimensional stacked history (projected on the $x$-$y$ plane) of all ionization produced throughout the exposure.
The data used for this analysis were acquired with a 1$\times$100 binning, where each pixel in the serial register collects the charge of 100 pixels in the corresponding column before the charge is moved in the $x$ direction and the serial register is read out. In this acquisition mode, each pixel in the image is effectively 15\,$\times$\,1500\,\um$^{2}$ in size. Since readout noise is introduced each time the charge is measured, a better signal-to-noise ratio in the measurement of the charge  is achieved by binning.  For details on the acquisition modes of DAMIC CCDs see Ref.~\cite{Aguilar-Arevalo:2016ndq}.

Nine exposures of 0.695\,d each were acquired with images 4622$\times$60 pixels in size.
The device was brought into inversion before every exposure to suppress surface dark current~\cite{janesick2001scientific}. 
The CCD data are contained in a 4116$\times$42-pixel segment of the image, corresponding to the physical size of the device, with the remaining regions constituting the ``overscan," where the CCD was clocked in both $x$ and $y$ directions beyond its active region to obtain measurements of zero charge with the primary output node.
Images were also acquired with the secondary CCD output node and with the other three CCDs installed in the DAMIC box.
Since the readout of all images is synchronized by the clocking, the noise images by the second output nodes of all CCDs allow for the identification and suppression of correlated electronic noise of the detector's readout chain.

The output of the CCD readout chain is recorded in analog-to-digital converter units (ADU) proportional to the number of charge carriers placed in the CCD's output node.
The linear constant $\alpha$ relating the pixel value to the number of charge carriers was calibrated before deployment with x-ray lines of known energy to be $\alpha$$=$0.0727\,$e^-$/ADU.
The linearity of the CCD output was confirmed within $\pm$10\% for signals as small as 2\,$e^-$ using optical photons~\cite{Aguilar-Arevalo:2016ndq}.

The image processing started with the subtraction of the pedestal (the constant offset of the pixel values introduced at the time of readout) from every pixel.
The pedestal was estimated independently for each row as the mean value of the pixels in the $x$ overscan.
To exclude a slight pedestal transient at the beginning of every row, the analysis was limited to the last 2500 columns of the image, for which the pixel values along rows in the $y$ overscan were found to be constant within statistical uncertainty.
To remove correlated readout noise, from every pixel we subtracted a linear combination of the pixel values in the corresponding four noise images, with the coefficients determined as to minimize the variance of the pixel noise.
Following this procedure, the noise in the images was estimated from the pixel values of the $y$ overscan to be \spix$=$1.9\,$e^-$.

We calculated the median and median absolute deviation (MAD) of every pixel over 114 images from a previous higher-temperature data set dedicated to background studies.
These quantities were used to construct a mask, which excludes localized dark current spikes due to defects in the silicon lattice~\cite{janesick2001scientific}.
These were identified as pixels that either deviate more than 3\,MAD from the median in at least 50\% of the images or have a median or MAD that is an outlier when compared to the distributions of these variables for all pixels.
This selection removed 0.17\% of the pixels.

This analysis considered all pixels with values up to 6\,\spix , including pixels that do not collect any charge and whose values arise solely from readout noise.
Thus, clusters of contiguous pixels with signal larger than 6\,\spix$=$11.4\,$e^-$ 
were masked from the image.
Pixels that were less than 4 pixels to the right or less than 200 pixels to the left of every cluster, i.e., within the 200 subsequent pixel readouts, were also masked.
This requirement rejected pixels with stray charge due to CCD charge transfer inefficiencies, which may happen when a high energy interaction results in a large number of charge carriers in the serial register.
Because of the low event rate from radioactive backgrounds ($\sim$1\,\pgd ), only 0.95\% of the pixels were removed by this procedure.

\begin{figure}[b!]
\centering
\includegraphics[width=0.482\textwidth]{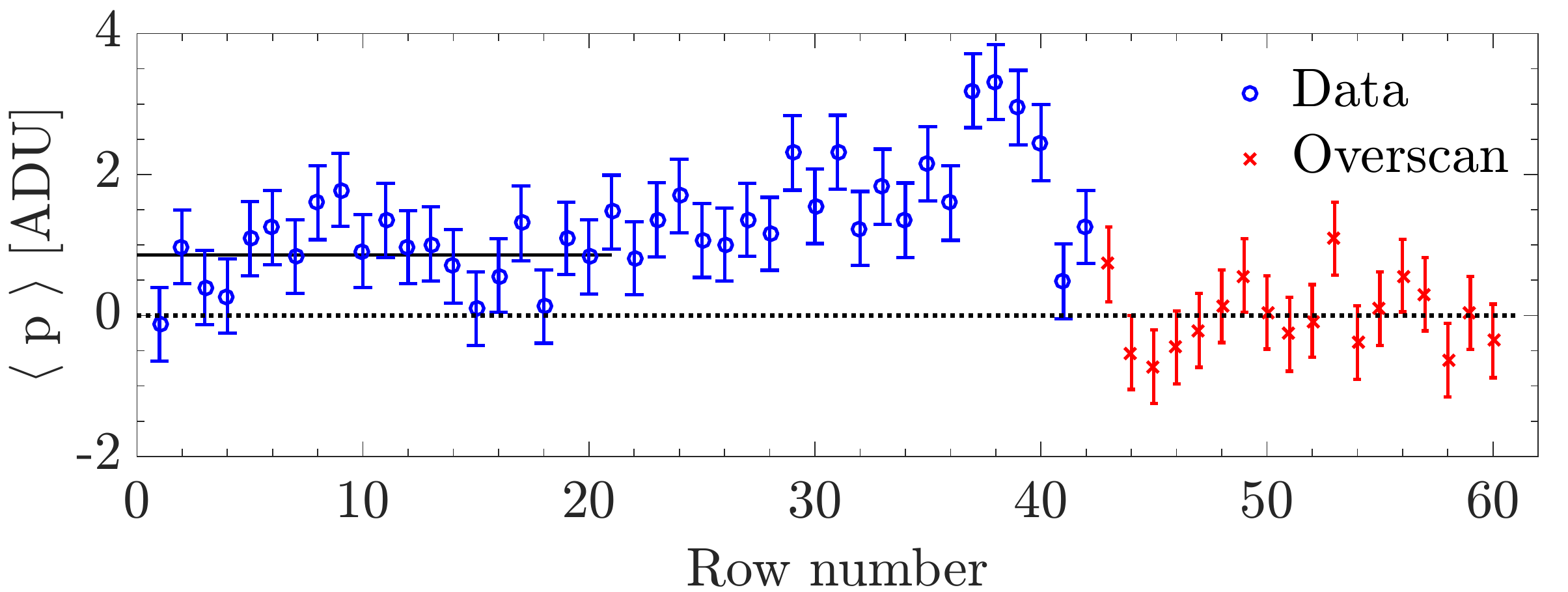}
\caption{Mean of the pixel values in each row over the nine images \pav , as a function of row number. The first 42 rows contain the CCD data, while the remaining 18 rows constitute the $y$ overscan. The dotted line shows \pav $=$0, while the solid line presents the level of leakage charge that corresponds to $\lambda$$=$4.0\,$e^-$\,mm$^{-2}$\,d$^{-1}$.}
\label{fig:dcrow}
\end{figure}

Figure~\ref{fig:dcrow} shows the mean value of pixels in each row over the nine images \pav\ after the image processing and pixel selection described above.
Rows $\ge$43 correspond to the $y$ overscan, for which \pav\ is consistent with zero.
The first 42 rows of the image contain the CCD data, for which an offset is clearly visible due to charge collected by the pixels. Hidden-photon absorption would produce charge uniformly distributed throughout the pixel array.
The higher values of collected charge in rows 30--40 indicate the presence of nonuniform sources of leakage current, e.g., optical or near-infrared photons inside the vessel or dark current exacerbated by mechanical stress of the silicon lattice.
The same pattern is more clearly observed in the other CCDs, with higher leakage current, installed in the DAMIC box, for which the increase is evident starting at row 22, suggesting that the charge distribution is spatially uniform only in the bottom half of the devices, i.e., rows 1--21.

\begin{figure}[b!]
\centering
\includegraphics[width=0.48\textwidth]{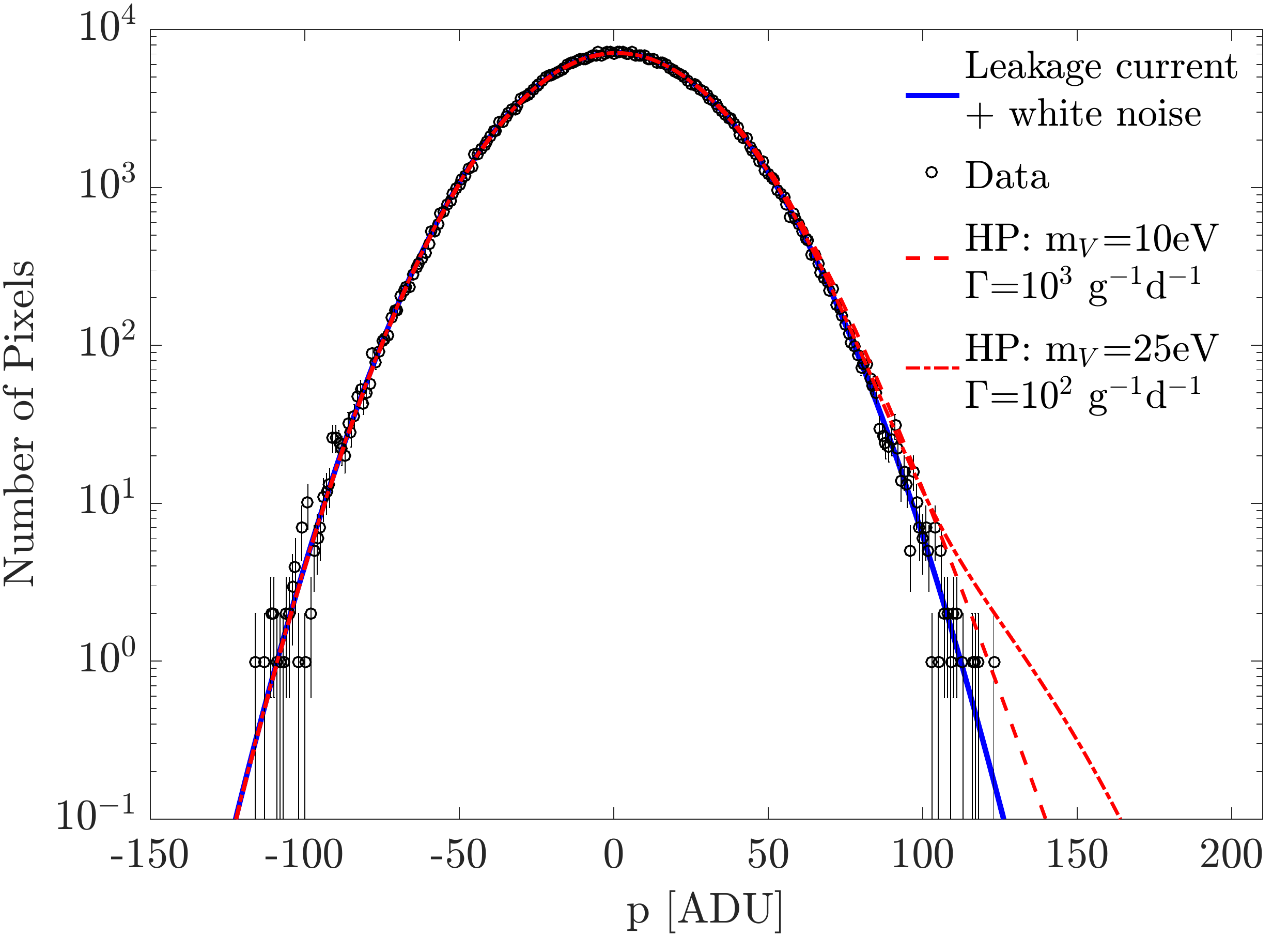}
\caption{Distribution of the pixel values $p$ considered for this analysis (markers). The solid line shows the best-fit result with the null hypothesis, i.e., only pixel white noise and a constant leakage current source across the device. The $p$ value is 0.78. The dashed (dot-dashed) line shows the result after including a fixed contribution from hidden photons with masses $m_V$$=$10\,eV (25\,eV) and an absorption rate of $\Gamma$$=$\powero{3}\,\pgd\ (\powero{2}\,\pgd ).}
\label{fig:fit}
\end{figure}

Thus, we consider rows 1--21 to place upper limits on the possible contribution to the collected charge from hidden photon absorption. This corresponds to $N$$=$\powert{5}{4.68} unmasked pixels over the nine images, equivalent to an exposure of 11.5\,g\,d. The distribution of pixel values $p$ (Fig.~\ref{fig:fit}) can be parametrized as
\begin{equation*} 
f(p) = N\sum_{n=0}^\infty F(n|\lambda,\Gamma,m_V) \textrm{Gaus}(\alpha p | n-\mu_0, \spix),
\end{equation*}
where $n$ is the number of charge carriers collected by a pixel, $F$ is their relative frequency, which depends on the leakage current per unit area $\lambda$ and the hidden-photon absorption rate $\Gamma$, and the Gaussian function describes the pixel white noise with mean $n$ and standard deviation $\sigma_{\rm pix}$. An offset $\mu_0$ that could remain because of the statistical uncertainty in the subtraction of the pedestal was included in the function.

In the absence of charge from hidden-photon absorption, i.e., for the case of the ``null" hypothesis, $F$ reduces to the contribution from leakage current:
\begin{equation}
F(n|\lambda,0,m_V) = \textrm{Poisson}(n|\mathcal{E}\lambda),
\label{eq:lc}
\end{equation}
modeled as a Poisson distribution under the assumption of uncorrelated production of charge carriers uniformly distributed across the selected region of the CCD.
The mean leakage charge collected by a pixel is proportional to the image single pixel exposure of $\mathcal{E}$$=$0.0156\,mm$^2$\,d.

To obtain the contribution to $F$ from the charge generated by hidden photon absorption we rely on Monte Carlo simulations.
For a given hidden photon mass $m_V$, we simulated a number of interactions drawn from a Poisson distribution with mean $\Gamma\mathcal{E}\rho N/9$, where $\rho=$1.57\,mg\,mm$^{-2}$ is the mass density per unit area of the CCD.
The spatial position of the hidden-photon absorption was uniformly distributed in the selected volume of the CCD.
For each simulated hidden-photon absorption, we generated the number of charge carriers as for the photoelectric absorption of a photon with energy $m_Vc^2$, using the probability distributions from Ref.~\cite{PhysRevB.22.5565, *scholze1998}.
We then distributed the carriers on the pixel array according to the charge diffusion model for the CCD, described and validated in Ref.~\cite{Aguilar-Arevalo:2016ndq}.
A histogram was made of the contents of all pixels in the simulated pixel array, and the simulation was repeated 100 times to obtain a numerical distribution of $F(n|0, \Gamma, m_V)$.
This function was then convolved with Eq.~(\ref{eq:lc}) to obtain $F(n|\lambda, \Gamma, m_V)$.

We first performed a likelihood fit to the data with the null hypothesis, with \spix , $\lambda$, and $\mu_0$ as free parameters.
Two penalty terms were added in the log-likelihood definition to include in the fit the prior knowledge of the values of \spix\ and $\mu_0$.
The value of \spix\ was constrained to the result from a fit to the pixels in the $y$ overscan, while $\mu_0$ was constrained within the statistical uncertainty in the pedestal subtraction.
The best-fit values were \spix$=$1.889$\pm$0.002\,$e^-$, $\lambda$$=$4.0$\pm$0.4\,$e^-$\,mm$^{-2}$\,d$^{-1}$, and $\mu_0$$=$0.010$\pm$0.005\,$e^-$, with a $p$ value of 0.78.

\begin{figure}[t!]
\centering
\includegraphics[width=0.478\textwidth]{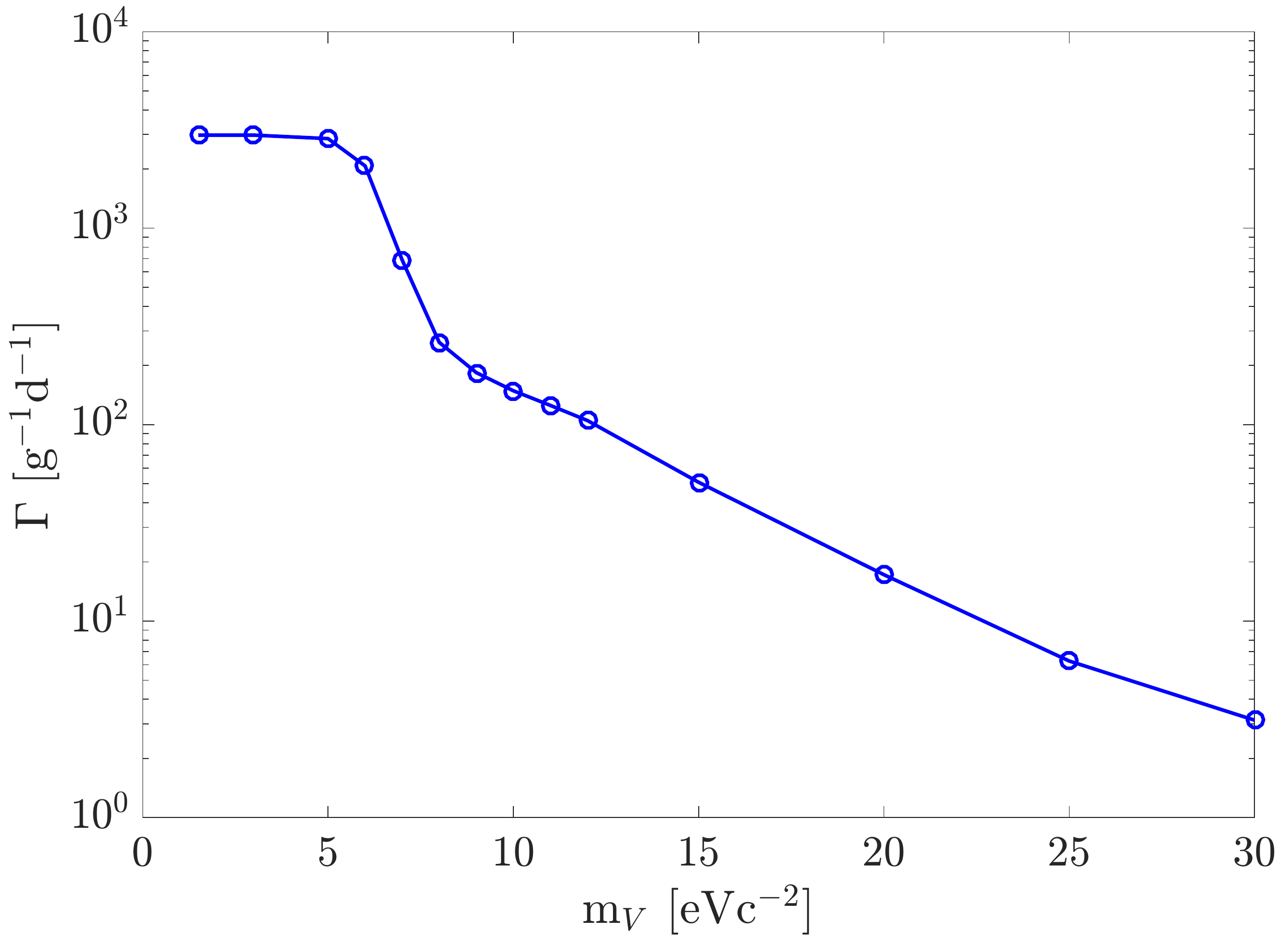}
\caption{Upper limits (90\% C.L.) on the hidden-photon absorption rate $\Gamma$ as a function of hidden-photon mass $m_V$ obtained from the likelihood fit described in the text.} \label{fig:rate}
\end{figure}

To explore the hidden-photon signal, we performed a scan for values of $m_V$ and $\Gamma$.
For each value of $m_V$, we increased the value of $\Gamma$ in discrete steps starting from zero.
At every step, the likelihood fit was performed where the parameters $m_V$ and $\Gamma$ were fixed and \spix , $\lambda$, and $\mu_0$ were free.
The minimum negative log-likelihood was registered and a likelihood profile was constructed for every $m_V$ as a function of $\Gamma$.
There was no statistical significance for a hidden-photon signal at any $m_V$.
The 90\% C.L. upper limit on $\Gamma$ was thus obtained from the likelihood profile using a likelihood-ratio test.
Figure~\ref{fig:rate} presents the results as a function of $m_V$ from 1.2 to 30\,\evm.

Below 5\,\evm , hidden-photon absorptions produce only one charge carrier, leading to a current source that would be indistinguishable from leakage current, and an upper limit on the absorption rate at the same level as the leakage current.
At higher $m_V$, the multiplicity in the number of carriers produced per absorption increases, leading to pixels that collect significantly more carriers than would be expected from leakage current.
This leads to a longer tail on the right-hand side of the pixel distribution, and consequently to a stronger upper limit on $\Gamma$.
To illustrate this, Fig.~\ref{fig:fit} shows the best-fit results with fixed parameters $m_V$$=$10\,eV and $\Gamma$$=$\powero{3}\,\pgd , and $m_V$$=$25\,eV and $\Gamma$$=$\powero{2}\,\pgd.

\begin{figure}[t!]
\centering
\includegraphics[width=0.48\textwidth]{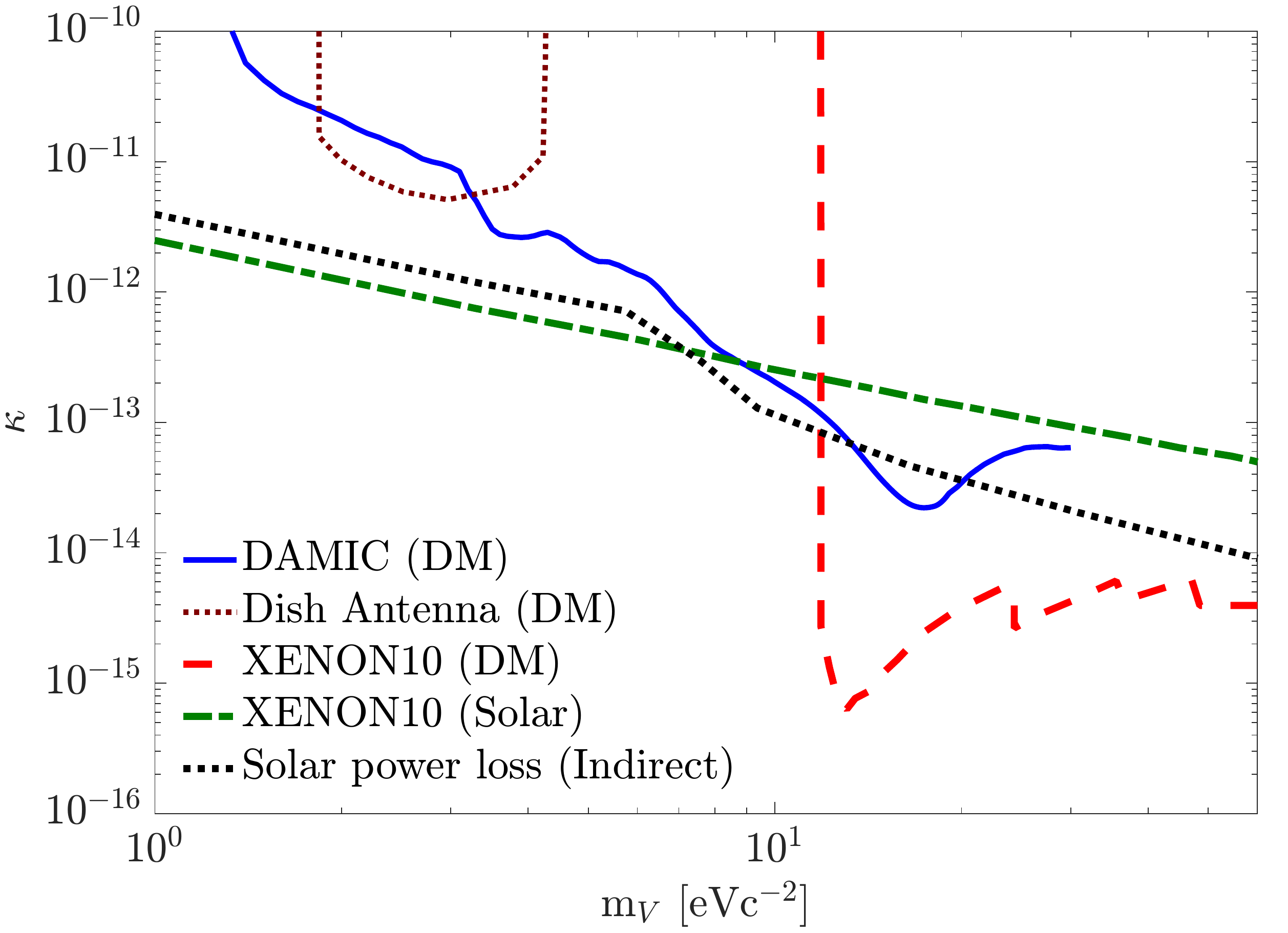}
\caption{Exclusion plot (90\% C.L.) for the hidden-photon kinetic mixing $\kappa$ as a function of hidden photon mass $m_V$ from the dark matter search presented in this Letter (solid line). The exclusion limits from other direct searches for hidden-photon dark matter in the galactic halo with a dish antenna (thin dotted line)~\cite{Suzuki:2015sza} and with the XENON10 experiment (dashed line)~\cite{Bloch:2016sjj}  are shown for comparison. A limit from a direct search with the XENON10 experiment for hidden photons radiated by the Sun (dot-dashed line)~\cite{Bloch:2016sjj} and an indirect constraint from the upper limit of the power lost by the Sun into invisible radiation (thick dotted line)~\cite{An:2013yfc, *Redondo:2013lna} are also presented.} \label{fig:limit}
\end{figure}

The absorption rate $\Gamma$ is related to the hidden-photon kinetic mixing $\kappa$ through $\kappa_{\rm eff}$ according to Eq.~(\ref{eq:rate}).
We use this relation to translate the upper limit on $\Gamma$ for a given $m_V$ to the corresponding upper limit on $\kappa$.
Following Ref.~\cite{Hochberg:2016sqx}, we compute the polarization tensor using the complex index of refraction in silicon, estimated at the detector operating temperature of 105\,K by extrapolating the values given in Ref.~\cite{Edwards1997547} using the empirical parametrization from Ref.~\cite{Rajkanan1979793}.
The results are shown in Fig.~\ref{fig:limit}.

Several sources of systematic uncertainty were investigated. The largest effect arises from the uncertainty in the linearity of the CCD output signal, which we estimated by varying $\alpha$ by $\pm$10\%, resulting in changes in the upper limit of $\Gamma$ ranging from 10\% for $m_V$$<$5\,\evm\ up to a factor of 2 for $m_V$$=$30\,\evm .
We repeated the analysis for different selected regions of the CCD, considering rows 1--18 and 1--24, or the last 2200 columns of the image.
This led to changes in the upper limits of $\Gamma$ especially at higher masses, with up to a 50\% increase at $m_V$$=$30\,\evm .
We confirmed the absence of pixels with values from 6 to 8\,\spix ; thus, the result is insensitive to the upper bound on the pixel values.
Finally, varying the temperature by $\pm$10\,K had a $<$5\% impact on the upper limits of $\kappa$.

The exclusion limits presented in this Letter are the most stringent direct-detection constraints on hidden-photon dark matter in the galactic halo with masses 3--12\,\evm .
The sensitivity of the experiment in terms of the kinetic mixing parameter $\kappa$ is approaching that of searches for hidden-photon emission by the Sun, offering a complementary technique for their detection.
Continued identification and mitigation of dark current and light sources in DAMIC will improve the sensitivity, making CCDs promising direct probes for hidden-photon dark matter with eV-scale masses.
In addition, this work characterizes the noise sources of DAMIC and demonstrates the sensitivity of the experiment to interactions that produce as little as a single electron, corresponding to ionization signals as small as 1.2\,eV.

We thank Tongyan Lin for motivating discussions on hidden-photon dark matter. We are grateful to SNOLAB and its staff for support through underground space, logistical and technical services. SNOLAB operations are supported by the Canada Foundation for Innovation and the Province of Ontario Ministry of Research and Innovation, with underground access provided by Vale at the Creighton mine site.
We acknowledge financial support from the following agencies and organizations: Kavli Institute for Cosmological Physics at the University of Chicago through Grants No. NSF PHY-1125897 and No. PHY-1506208 and an endowment from the Kavli Foundation; Fermi National Accelerator Laboratory (Contract No. DE-AC02-07CH11359); Institut Lagrange de Paris Laboratoire d'Excellence (under Reference No. ANR-10-LABX-63) supported by French state funds managed by the Agence Nationale de la Recherche within the Investissements d'Avenir program under Reference No. ANR-11-IDEX-0004-02; Swiss National Science Foundation through Grant No. 200021\_153654 and via the Swiss Canton of Zurich; Mexico's Consejo Nacional de Ciencia y Tecnolog\'{i}a (Grant No. 240666) and  Direcci\'{o}n General de Asuntos del Personal Acad\'{e}mico - Universidad Nacional Aut\'{o}noma de M\'{e}xico (Programa de Apoyo a Proyectos de Investigaci\'{o}n e Innovaci\'{o}n Tecnol\'{o}gica Grants No. IB100413 and No. IN112213); Brazil's Coordena\c{c}\~{a}o de Aperfei\c{c}oamento de Pessoal de N\'{\i}vel Superior, Conselho Nacional de Desenvolvimento Cient\'{\i}fico e Tecnol\'{o}gico, and  Funda\c{c}\~{a}o de Amparo \`{a} Pesquisa do Estado de Rio de Janeiro.

\bibliography{myrefs.bib}

\end{document}